\documentclass[12pt]{iopart}
\usepackage{iopams}
\usepackage[dvips]{graphicx}
\usepackage{amsfonts}
\usepackage{psfrag}

\newcommand{\bea}{\begin{eqnarray}}
\newcommand{\eea}{\end{eqnarray}}
\newcommand{\beq}{\begin{equation}}
\newcommand{\eeq}{\end{equation}}
\newcommand{\nn}{\nonumber}
\newcommand{\eqref}[1]{(\ref{#1})}

\newcommand{\mubar}{{\overline \mu}}

\newcommand{\half}{{\frac{1}{2}}}

\newcommand{\third}{{\frac{1}{3}}}

\newcommand{\abs}[1]{\vert #1\vert}
\newcommand{\avg}[1]{\left\langle #1 \right\rangle_{\mu}}
\newcommand{\avgp}[1]{\left\langle #1 \right\rangle_{\mubar}}
\newcommand{\avgA}[1]{\left\langle #1 \right\rangle_{\mu^A}}
\newcommand{\avgB}[1]{\left\langle #1 \right\rangle_{\mu^B}}
\newcommand{\ep}{\epsilon}

\begin{document}

\title{Biased random walks on random combs }
\author{Tanya M Elliott and John F Wheater}
\address{Department of Physics, University of Oxford \\
Rudolf Peierls Centre for Theoretical Physics,\\
1 Keble Road,\\
 Oxford OX1 3NP, UK}
\eads{\mailto{t.elliott1@physics.ox.ac.uk}, \mailto{j.wheater@physics.ox.ac.uk}}
\begin{abstract}
We develop rigorous, analytic techniques to study the behaviour of biased random walks on combs.  This enables us to calculate exactly the spectral dimension of random comb ensembles for any bias scenario in the teeth or spine.  Two specific examples of random comb ensembles are discussed; the random comb with nonzero probability of an infinitely long tooth at each vertex on the spine and the random comb with a power law distribution of tooth lengths.  We also analyze transport properties along the spine for these probability measures.  
\end{abstract}


\section{Introduction\label{INTRO}}
The behaviour of random walks on random combs is of interest from a number of points of view.  
Condensed matter physicists have studied such structures because they serve as a model for diffusion in more complicated 
fractals and percolation clusters \cite{one,two,three,halv}.  In the context of quantum gravity, 
random combs are a tractable example of a random manifold ensemble and understanding their 
geometric properties can provide insight into higher dimensional problems \cite{four,dh,specdim1}.  Most of the literature concerns approximate analytical techniques and numerical solutions, although there are exact calculations of leading order behaviour in some cases \cite{Pot1}.  To this end, it is desirable to have rigorous methods for determining the geometric quantities of interest and that is the purpose of this paper.  

One such quantity is the dimensionality of the ensemble.  On a sufficiently smooth manifold all definitions 
of dimension will agree, but for fractal geometries like random combs this is not necessarily 
true.  The spectral dimension is defined to be \(d_s\) provided the ensemble average 
probability of a random walker being back at the origin at time \(t\), takes the asymptotic 
form $t^{-d_s/2}$.  This concept of dimension does not in general agree with the Hausdorff dimension \(d_H\), 
which is defined when the expectation value of the volume enclosed within a geodesic 
distance \(R\) from a marked point scales like $R^{d_H}$ as \(R \to \infty\).

We know that for diffusion on regular structures the mean square displacement at large times is proportional 
to \(t\), but for a fractal substrate there is anomalous diffusion and the mean square displacement behaves like  
$t^{2/d_w}$, where \(d_w\) represents the fractal dimension of the walk and depends sensitively on the nature of the random structure.  

\emph{Biased} random walks on combs have also been studied in connection with disordered 
materials, since such a system is a paradigm for diffusion on fractal structures in the 
presence of an applied field \cite{six,eight}.  As we discuss later there are several different bias regimes.  
Topological bias, where at every vertex in the comb there is an increased probability of moving away from the origin was first studied for a 
random comb with a power-law distribution of tooth lengths in \cite{seven}.  Other works have discussed the effects of bias away from the origin only in the teeth \cite{Pot3} and only in the spine \cite{Pot2}.  The effect of going into the teeth can be viewed as creating a waiting time for the walk along the spine; the distribution of the waiting time depends on both the bias and the length of the teeth and the outcome is the result of subtle interplay between the two.  

In \cite{comb} some new, rigorous techniques were developed to study random 
walks on combs.  This enabled an exact, but very simple calculation of the spectral dimension of random combs. 
The principal idea is to split both random combs and random walks into subsets 
that give either strictly controllable or exponentially decaying contributions to the 
calculation of physical characteristics.  These methods were later reinforced to prove that 
the spectral dimension of generic infinite tree ensembles is 4/3 \cite{trees,trees2}.  In this paper we use and extend the techniques of \cite{comb} to deal with biased walks on combs.  Some of our results are new; some qualify statements made in the literature; and some merely confirm results already derived by other, usually less rigorous, methods.

The random combs, the bias scenario, some useful generating functions and the critical exponents are 
defined in the next section.  In Section 3 we introduce some deterministic combs, discuss 
general properties of the generating functions and establish bounds that will be instrumental 
when studying random ensembles.  Section 4 looks at regions of bias where the large time 
behaviour is independent of the comb ensemble or simply dependent on the expectation value of 
the first return generating function in the teeth.  In Section 5 we compute the spectral 
dimension in regions of bias where it is influenced by the probability measure on the teeth.  
Two specific cases are considered: the random comb with nonzero probability of an infinitely 
long tooth at each vertex on the spine and the random comb with a power law distribution of 
tooth lengths.  Section 6 examines transport properties along the spine for these same 
probability measures and in the final section we review the main results, compare with the literature and discuss their significance.  
Some exact calculations and proofs omitted from the main text are outlined in the appendices.  

\section{Definitions\label{DEFS}}

Wherever possible we use the definitions and notation of \cite{comb}; we repeat them here for the reader's convenience but mostly refer back to \cite{comb} for proofs and derived properties.
\subsection{Random combs}
Let $N_\infty$ denote the nonnegative integers regarded as a graph so that $n$
has the neighbours $n\pm 1$ except for $0$ which only has $1$ as a neighbour.
Let $N_\ell$ be the integers $0,1,\ldots ,\ell$ regarded as a graph so that
each integer $n\in N_\ell$ has two neighbours $n\pm 1$ except for $0$ and $\ell$
which only have one neighbour, $1$ and $\ell-1$, respectively.  A comb 
$C$ is an
\begin{figure}[h!]
\begin{center}
\includegraphics{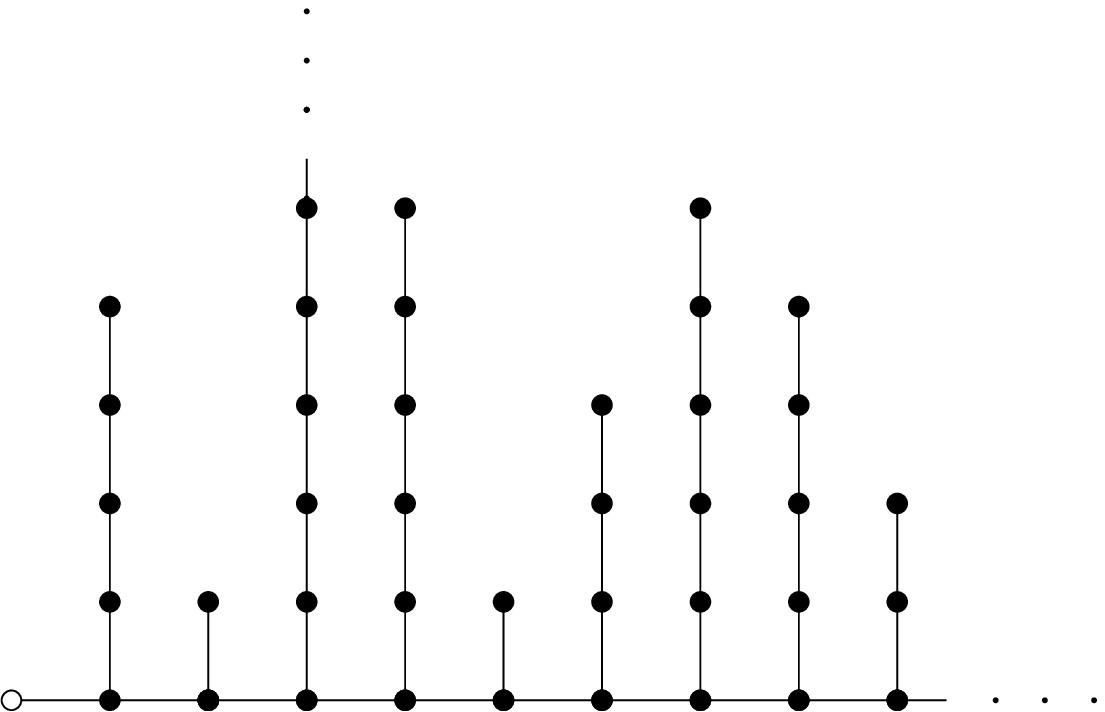}
\caption{A comb.}
\label{fig1}
\end{center}
\end{figure}
infinite rooted tree-graph with a special subgraph $S$ called the spine
which is isomorphic to $N_\infty$ with the root, which we denote $r$, at $n=0$.  At each vertex of
$S$, except the root $r$,  there is  attached by their endpoint $0$ one of the graphs $N_\ell$ or
$N_\infty$.   The linear graphs
attached to the spine are called the teeth of the comb, see figure 1. 
  We will denote by $T_n$ the tooth attached to
the vertex $n$ on $S$, and by $C_k$ the comb obtained by removing the links 
$(0,1),\ldots ,(k-1,k)$, the teeth $T_1,\ldots ,T_k$ and relabelling the 
remaining vertices on the spine in the obvious way. An arbitrary comb is specified by a list of its teeth $\{T_1,\ldots\}$ and \(\abs{T_k} \) denotes the length of the tooth.    Note that we have excluded the possibility of a tooth of zero length. This is   for technical convenience in what follows  and can be relaxed \cite{TanyaThesis}.

In this paper we are interested in random combs for which the length $\ell$  of each  tooth is identically and independently distributed with probability $\mu_\ell$. This induces a probability measure $\mu$ on the positive integers and expectation values with respect to this measure will be denoted $\langle \cdot\rangle_\mu$. In particular we will consider the two measures
\bea 
\mu^A_\ell&=&\cases{p,&$\ell=\infty,$\\1-p,& $\ell=1,$\\0,&\rm{otherwise;}}\nn\\
\mu^B_\ell&=&\frac{C_a}{\ell^a},\quad a>1.\eea
However, the results proved for $\mu^B$ apply to any measure with the same behaviour at large $\ell$ and we note in passing that the methods used here will work for any distribution that is reasonably smooth, for example the exponential distribution.  The measure $\mu^B$ has been discussed quite extensively in the literature but $\mu^A$ has not.  

\subsection{Biased random walks}

We regard time as integer valued and consider a walker who makes one step on the graph for each unit time interval. If the walker is at the root or at the end-point of a tooth then she leaves with probability 1. If at any other vertex the probabilities are parametrized by two numbers $\epsilon_1$ and $\epsilon_2$ as shown in figure 2a and the allowed range of these parameters is shown in figure 2b. For walks in the teeth there is bias away from or towards the spine depending on whether $\epsilon_2$ is positive or negative; similarly a walk on the spine is biased away from or towards the root depending on whether $\epsilon_1$ is positive or negative. When there is no bias we say that the walk is `critical'; the fully critical case $\epsilon_1=\epsilon_2=0$ was covered in \cite{comb}.  The notation
\bea
b_- &=& 1-\epsilon_1-\epsilon_2, \nn \\
b_+ &=& 1+\epsilon_1-\epsilon_2, \nn \\
b_T &=& 1+2\epsilon_2,
\eea
will be used where applicable since these combinations appear often in our analysis.  We denote by $B, B',B_1, B_2$ etc constants which  depend on $\epsilon_1$ and $\epsilon_2$ and may vary from line to line but are positive and finite on the relevant range; other constants will be denoted  $c,c'$ etc.

\psfrag{1/2(1+2ep2)}{$\frac{1}{2}(1+2\epsilon_2)$}
\psfrag{1/2(1-2ep2)}{$\frac{1}{2}(1-2\epsilon_2)$}
\psfrag{1/3(1+2ep2)}{$\frac{1}{3}(1+2\epsilon_2)$}
\psfrag{1/3(1-ep1-ep2)}{$\frac{1}{3}(1-\epsilon_1-\epsilon_2)$}
\psfrag{1/3(1+ep1-ep2)}{$\frac{1}{3}(1+\epsilon_1-\epsilon_2)$}
\psfrag{ep1}{{\large$\epsilon_1$}}	
\psfrag{nep2}{$\epsilon_2$}
\psfrag{nep1}{$\epsilon_1$}	
\psfrag{ep2}{{\large$\epsilon_2$}}
\psfrag{1/2}{{\large$\frac{1}{2}$}}	
\psfrag{-1/2}{-{\large$\frac{1}{2}$}}	

\begin{figure}[h!]
\begin{center}
\centerline{\hbox{\scalebox{0.8}{\includegraphics{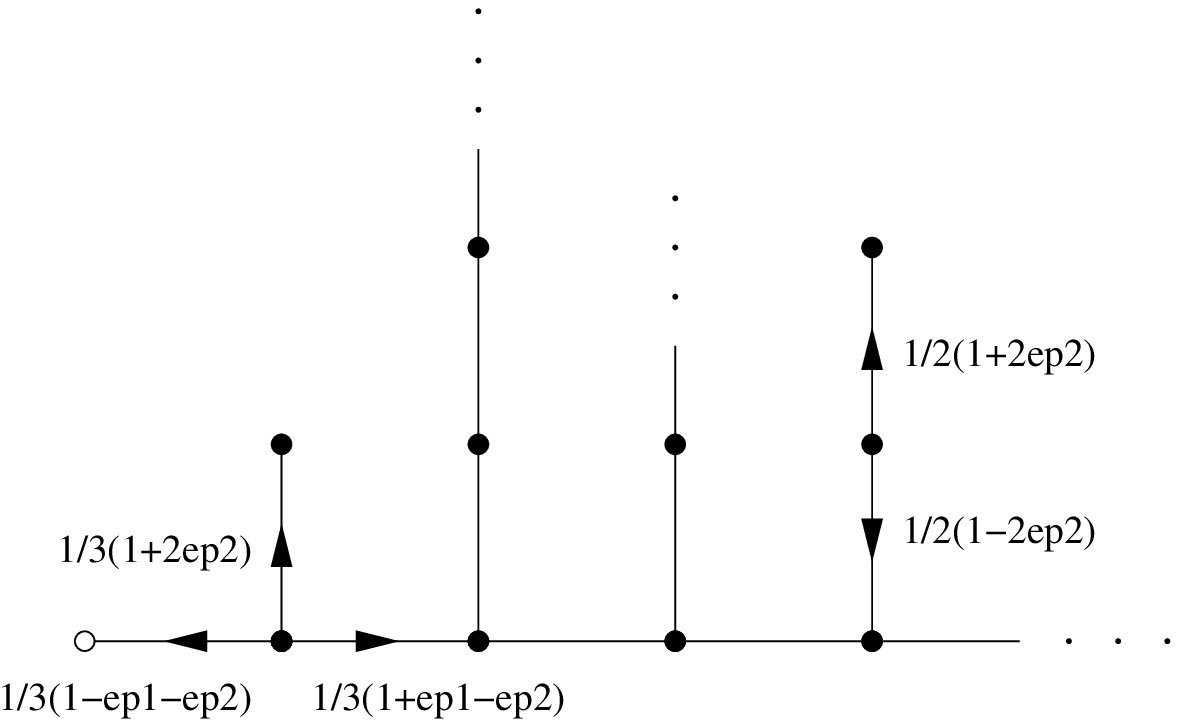}}}}
\centerline{\hbox{(a)}}
\vspace{10pt}
\centerline{\hbox{\scalebox{0.8}{\includegraphics{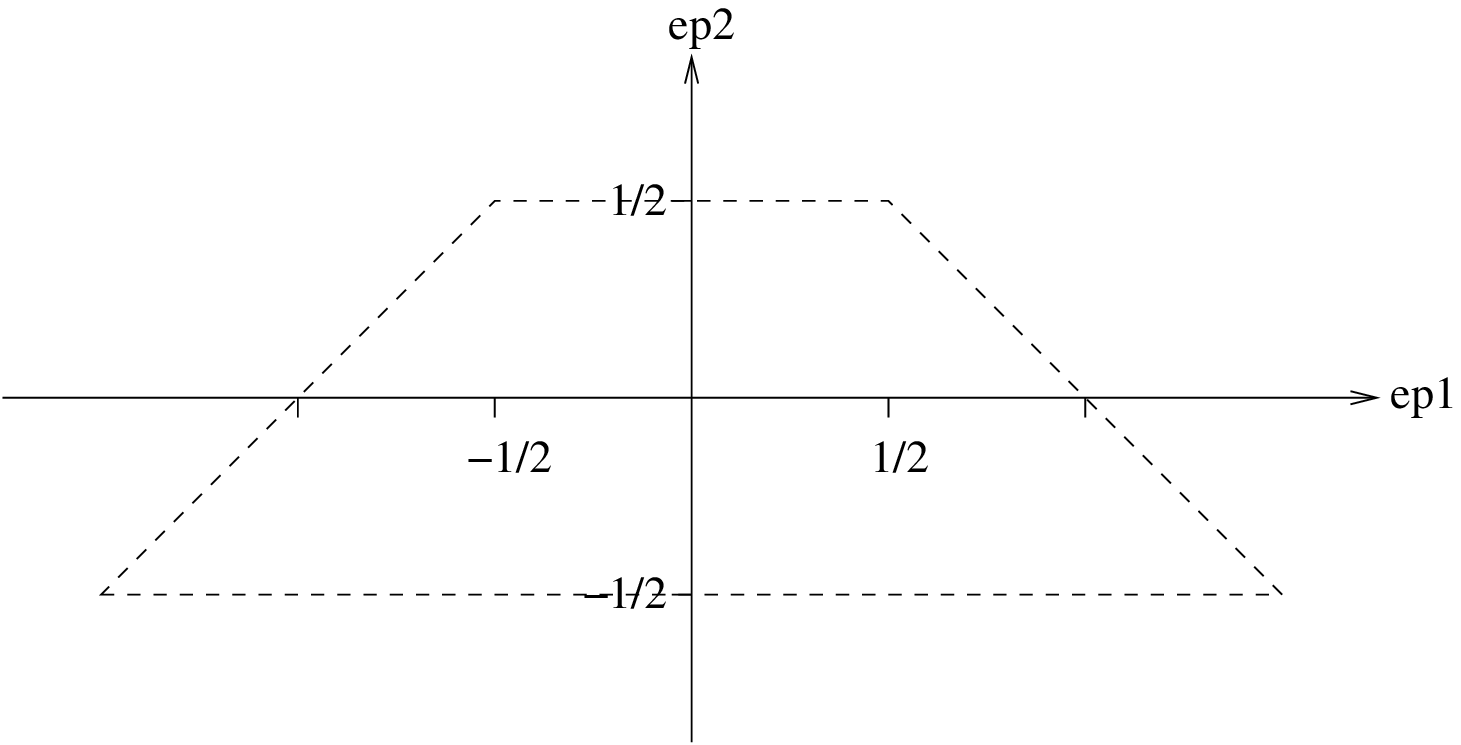}}}}
\centerline{\hbox{(b)}}
\caption{Bias parameterisation.}
\label{fig2}
\end{center}
\end{figure}

The generating function for  the probability $p_C(t)$ that the walker on $C$ is back at the root at time $t$ having left it at $t=0$  is defined by
\beq Q_C(x) = \sum_{t=0}^\infty (1-x)^{t/2}p_C(t).\eeq
Letting $\omega$ be a walk on $C$ starting at $r$, $\omega(t)$ the vertex where the walker is to be found at time $t$, and $\rho_{\omega(t)}$ the probability for the walker to step from $\omega(t)$ to $\omega(t+1)$, we have
\beq Q_C(x) = \sum_{\omega: r\to r}(1-x)^{\half\vert\omega\vert}\prod_{t=0}^{\vert\omega\vert-1}\rho_{\omega(t)}.\eeq
A similar relation gives the generating function for probabilities for first return to  the root, $P_C(x)$, except that the trivial walk of duration 0 is excluded. The two functions are related by
\beq Q_C(x) = \frac{1}{1-P_C(x)},\label{QPreln}\eeq
and it is straightforward to show that $P_C(x)$ satisfies the recurrence relation
\beq P_C(x) = \frac{(1-x)b_-}{3-b_+P_{C_1}(x)-b_TP_{T_1}(x)}.\label{recur}\eeq
Note that $P_C(x)$ and $Q_C(x)$ depend upon $\epsilon_1$ and $\epsilon_2$; to avoid clutter we will normally suppress this dependence but if necessary it will appear as superscripts.  It is important for what follows that $Q_C$ is a convex function of $P_C$ which is itself a convex function of $P_{T_1},...P_{T_k},P_{C_k}$ for any $k>0$.  

For an ensemble of combs, we will denote the expectation values of the generating functions for return and first return probabilities as
\bea
Q(x) &=& \avg{Q_C(x)} \nn \\
P(x) &=& \avg{P_C(x)}.
\eea
We will say that $g(x) \sim f(x)$ if there exist positive constants $c$, $c'$, $\sigma$, $\sigma'$ and $x_0$ such that 
\beq
c \; f(x) \exp \left(-\sigma (\log \abs{f(x)})^{1/a} \right) < g(x) < c' \; f(x) \abs{\log f(x)}^{\sigma'}
\eeq
for $0 < x \le x_0$.  The tactic of this paper is to prove bounds of this form for the generating functions; in almost all cases our results are in fact a little stronger having $\sigma=\sigma'=0$ when we will say that $g(x) \approx f(x)$.  

The random walk on $C$ is recurrent if $P_C(0)=1$ in which case we define the exponent $\beta$ through
\beq
1-P_C(x) \sim x^\beta.
\eeq
If $\beta$ is an integer then we expect logarithmic corrections and define $\tilde{\beta}$ if
\beq
1-P_C(x) \approx x^\beta \abs{\log x}^{-\tilde{\beta}}.
\eeq
It follows that $Q_C(x)$ diverges as $x \to 0$ and we define \(\alpha\) by
\beq
Q_C(x) \sim x^{-\alpha},
\eeq
and if \(\alpha\) is an integer, \(\tilde{\alpha}\) when
\beq
Q_C(x) \approx x^{-\alpha} \abs{\log x}^{\tilde{\alpha}}.
\eeq
If \(P_C(0) <1\) then the random walk is non-recurrent, or transient, and \(Q_C(x)\) is finite as \(x \to 0\).  Then if, as \(x \to 0\), the first \(k-1\) derivatives of \(Q_C(x)\) are finite but the \(k\)th derivative diverges we define the exponent \(\alpha_k\) by
\beq
Q_C^{(k)}(x) \sim x^{-\alpha_k},
\eeq
and if \(\alpha_k\) is an integer, \(\tilde{\alpha}_k\) when
\beq \label{dsdefn}
Q_C^{(k)}(x) \approx x^{-\alpha_k} \abs{\log x}^{\tilde{\alpha}_k}.
\eeq

In considering the ensemble of combs \(\mu\), we define all these exponents in exactly the same way simply replacing \(P_C(x)\) with \(\avg{P_C(x)}\) and so on.  Note that for a single recurrent comb \(\beta=\alpha\) but in an ensemble this is no longer necessarily the case; applying Jensen's inequality to \eqref{QPreln} we see that \(\beta \le \alpha\).  

If \(Q^{(k)}(x) \sim x^{-\alpha_k}\) then it is straightforward to show that
\beq
R^{k}(\lambda) = \sum_{t=0}^{\lambda^{-1}} t^k \; \avg{p_C(t)} \sim \lambda^{-\alpha_k}.
\eeq
It follows that if the sequence decays uniformly at large \(t\), which we do not prove, then it falls off as \(t^{\alpha_k-1-k}\).  Thus we define \(d_s=2(1+k-\alpha_k)\).  Similarly if \(Q(x) \approx \abs{\log x}^{\tilde{\alpha}}\) then \(R(\lambda) \approx \abs{\log \lambda}^{\tilde{\alpha}}\) and, again assuming uniformity, \(p(t)\) falls off as \(t^{-1} \; \abs{\log t}^{\tilde{\alpha}-1}\).

\subsection{Two-point functions}

Let $p^1_C(t;n)$ denote the probability that the walker on $C$,  having left $r$ at $t=0$ and not subsequently returned there, is at point $n$ on the spine at time $t$.  The corresponding generating function, which we will call the two-point function, is defined by
\beq G_C(x;n) = \sum_{t=0}^\infty (1-x)^{t/2}p^1_C(t;n).\eeq
Letting $\omega$ be a walk on $C$ starting at $r$ and ending at $n$ without returning to $r$ we have
\beq G_C(x;n) = \sum_{\omega: r\to n}(1-x)^{\half\vert\omega\vert}\prod_{t=0}^{\vert\omega\vert-1}\rho_{\omega(t)}.\eeq
Following the discussion in section 2.2 of \cite{comb} this leads us to the representation
\beq G_C(x;n) =\frac{3}{b_+(1-x)^{n/2}}\prod_{k=0}^{n-1}\frac {b_+}{b_-}P_{C_k}(x).\label{GCreln}\eeq

\subsection{The Heat kernel}
Let $K_C(t;n,\ell)$ denote the probability that the walker on $C$,  having left $r$ at $t=0$, is at point $\ell$ in tooth $T_n$ at time $t$. $K_C(t;n,\ell)$ satisfies the diffusion equation on $C$ so we call it the heat kernel. The probability that the walker has travelled a distance $n$ along the spine at time $t$ is given by
\beq K_C(t;n)=\sum_{\ell=0}^\infty K_C(t;n,\ell), \eeq
and has generating function
\beq H_C(x;n)=\sum_{t=0}^\infty (1-x)^{t/2} K_C(t;n).\eeq
$H_C(x;n)$ can be written as
\beq H_C(x;n)= \frac{G_C(x;n)}{1-P_C(x)}D_{\abs{ T_n}}(x),\label{heatk}\eeq
where
\beq D_{\ell}(x)=1+\frac{b_T}{3}\sum_{k=1}^\ell G_{N_\ell}(x;k),\eeq
and we define
\beq H(x;n)=\avg{H_C(x;n)}.\eeq
Note that, because $K_C(t;n)$ is a probability,
\beq \sum_{n=0}^\infty H(x;n) =\frac{1+\sqrt{1-x}}{x}.\label{unitarity}\eeq

The exponent $d_k$ is defined through the moments in $n$
\beq
\sum_{n=0}^{\infty} n^k \; H(x;n) \approx x^{-1-d_k},
\eeq
and in the case $d_k=0$ the exponent $\tilde{d}_k$ is defined when
\beq
\sum_{n=0}^{\infty} n^k \; H(x;n) \approx x^{-1}\; \abs{\log x}^{\tilde{d}_k}.
\eeq
If $\ep_1 \ge 0$ one can show that on any comb $\langle n \rangle_{\omega: \abs{\omega}=t}$ is a non-decreasing sequence and thus that there is some constant $T_0$ such that for $T > T_0$
\beq
\begin{array}{llllll}
\underline{c} \: \left(\frac{T}{\abs{\log T}} \right)^{d_1} & < & \avg{\langle n \rangle_{\omega: \abs{\omega}=T} + \langle n \rangle_{\omega: \abs{\omega}=T+1}} & < & \overline{c} \: T^{d_1}, & d_1 \ne 0 \nonumber \\
\underline{c} \phantom{\int} \abs{\log T}^{\tilde{d}_1} & < & \avg{\langle n \rangle_{\omega: \abs{\omega}=T} + \langle n \rangle_{\omega: \abs{\omega}=T+1}} & < & \overline{c} \: \abs{\log T}^{\tilde{d}_1}, & d_1=0.
\end{array}
\eeq
If $\ep_1<0$ (for which we always have $\tilde{d}_1=0$) then we have only the weaker result that for $T>T_0$
\beq
\underline{c} \: \left(\frac{T}{\abs{\log T}} \right)^{1+d_1}  <  \sum_{t=0}^{T} \avg{\langle n \rangle_{\omega: \abs{\omega}=t}} < \overline{c} \: T^{1+d_1}.
\eeq

\section{Basic properties}

\subsection{Results for simple regular combs}
The relation (\ref{recur}) can be used to compute the generating functions for a number of simple regular graphs which will be important in our subsequent analysis \cite{comb}. 
\begin{enumerate}
\item An infinitely long tooth, $N_\infty$:
\beq P_\infty(x)=\cases{1-x^\half  & {if $\epsilon_2=0$;}\\
                               \frac{1-2\abs{\epsilon_2}}{b_T}-\frac{x}{4\abs{\epsilon_2}}(1-2\epsilon_2) +O(x^2)& {otherwise.}}\label{Pinf}\eeq

\item A tooth of length $\ell$, $N_\ell$:
\beq P_\ell(x)= P_\infty(x)\frac{1+XY^{1-\ell}}{1+XY^{-\ell}}\label{Pell}\eeq
where
\beq  X=\frac{b_T(1-P_\infty(x))}{2-b_T(1+P_\infty(x))},\quad Y=\frac{2-b_TP_\infty(x)}{b_TP_\infty(x)}\,.\eeq

\item The comb $\sharp$ given by $\{T_k=N_1,\forall k\}$ has all teeth of length 1, and
\beq\fl P_\sharp(x)=\cases{1-B_1 x^\half+O(x) & {if $\epsilon_1=0$;}\\
                               \frac{1-\epsilon_2-\abs{\epsilon_1}}{b_+}-x\frac{ B_2}{\abs{\epsilon_1}} +O(x^2)& {otherwise.}}\label{Psharp}\eeq
Note that $\sharp$ is non-recurrent if $\epsilon_1>0$.
 It is also convenient to define $\ell\sharp$ to be $\{T_1=N_\ell,C_1=\sharp\}$.
\item The comb $*$ given by $\{T_k=N_\infty,\forall k\}$ has all teeth of length $\infty$ and is non-recurrent for $\epsilon_2>0$,
\beq\fl P_*(x)=\frac{1+\epsilon_2-\sqrt{4\epsilon_2+\epsilon_1^2}}{b_+}-x\frac{B_1}{\sqrt{4\epsilon_2+\epsilon_1^2}}
+O(x^2).\label{PstarA}\eeq
Otherwise
\beq\fl P_*(x)=\cases{
\frac{1-\abs{\epsilon_1}}{1+\epsilon_1}-\frac{B_2}{\abs{\epsilon_1}} x^\half +O(x)& {if $\epsilon_2=0$, $\epsilon_1\ne0$;}\\
                               \frac{1-\epsilon_2-\abs{\epsilon_1}}{b_+}-x\frac{ B_3}{\abs{\epsilon_1}} +O(x^2)& {if $\epsilon_2<0$, $\epsilon_1\ne0$;}\\
1-B_4 x^\half +O(x)& {if $\epsilon_2<0$, $\epsilon_1=0$.}}\label{PstarB}\eeq

\item The comb $\flat\ell$ given by $\{T_k=N_\ell,\forall k\}$ has all teeth of length $\ell$ and
\beq\fl P_{\flat\ell}(x)=\cases{
\frac{1-\abs{\epsilon_1}}{1+\epsilon_1}-\frac{B_1}{\abs{\epsilon_1}}(\ell+1+\abs{\epsilon_1}) x +O(x^2\ell^2)& {if $\epsilon_2=0$, $\epsilon_1\ne0$;}\\
                               \frac{1-\epsilon_2-\abs{\epsilon_1}}{b_+}-x\frac{ B_2}{\abs{\epsilon_1\epsilon_2}} +O(xY^{-\ell})& {if $\epsilon_2<0$, $\epsilon_1\ne0$;}\\
1-\frac{B_3}{\abs{\epsilon_2}} x^\half +O(x^\half Y^{-\ell})& {if $\epsilon_2<0$, $\epsilon_1=0$;}}\label{PflatA}\eeq
where, as \(x \to 0 \),
\beq Y \to \frac{1+2\abs{\epsilon_2}}{1-2\abs{\epsilon_2}}. \label{Y}\eeq
When \(\epsilon_2 > 0\) let \(\bar{\ell} = \lfloor \abs{\log x}/ \log Y \rfloor \), where $\lfloor z \rfloor$ denotes the integer below \(z\). For $\ell > 2\bar{\ell}$ the teeth are long enough that \(P_{\flat\ell}(x)\) behaves like (\ref{PstarA}).  For $\bar{\ell} < \ell \le 2\bar{\ell}$, \(P_{\flat\ell}(x)\) is non-recurrent with the leading power of $x$ being fractional.  For $\ell \le \bar{\ell}$
\beq\fl P_{\flat\ell <\bar{\ell}}(x)=\cases{
\frac{1-\epsilon_2-\abs{\epsilon_1}}{b_+}-x\frac{ B_4 Y^{\ell}}{\abs{\epsilon_1\epsilon_2}} +O(x)& {if $\epsilon_1\ne0$;}\\
1-\frac{B_5}{\sqrt{\epsilon_2}} x^\half Y^{\half \ell} +O(x^{\half}Y^{-\half\ell},xY^{-\ell}) & {if $\epsilon_1=0,$}}\label{PflatB}\eeq
where the notation $O(a,b)$ means $O(\max{(a,b)})$.  
\end{enumerate}

\subsection{General properties of the generating functions}

The generating functions for any comb satisfy three simple properties which can be derived from (\ref{recur}):
\begin{enumerate}
\item \emph{Monotonicity} The value of $P_C(x)$ decreases monotonically if the length of a tooth is increased.
\item \emph{Rearrangement} If the comb $C'$ is created from $C$ by swapping the adjacent teeth $T_k$ and $T_{k+1}$ then $P_{C'}(x) > P_{C}(x)$ if $\abs{T_{k+1}}<\abs{T_{k}}$.
\item \emph{Inheritance} If walks on $C_k$ or $T_k$  are non-recurrent for finite $k$ then walks on $C$ are non-recurrent.
\end{enumerate}
The proof of the first two follows that given in \cite{comb} for the special case $ \epsilon_2=\epsilon_1=0$. The third can be shown by assuming that either $P_{C_1}(0)<1$ or $P_{T_1}(0)<1$; it then follows immediately from (\ref{recur}) that $P_C(0)<1$ and the result follows by induction.

\subsection{Useful elementary bounds}

By monotonicity $G_C(x;n)$ is always bounded above by $G_{\sharp}(x;n)$ from which we get 
\beq G_C(x;n) < \frac{3}{b_+}\exp(-n\Lambda^{\epsilon_1,\epsilon_2}(x)),\label{Gbound}\eeq
where
\beq \Lambda^{\epsilon_1,\epsilon_2}(x) = \cases{x\,\frac{2+\ep_2}{2\ep_1} & {if $\epsilon_1>0$,}\\
                      x^\half\sqrt{\frac{2+\ep_2}{1-\ep_2}} &  {if $\epsilon_1=0$,}\\
                    \log\left(\frac{b_-}{b_+}\right) &    {if $\epsilon_1<0$.}}\eeq
Now let $P_C^{(N)}(x)$ denote the contribution to $P_C(x)$ from walks that reach beyond $n=N$ on the spine. It is straightforward to show using the arguments of section 2.5 of \cite{comb} that 
\beq P_C^{(N)}(x)\le \third b_- G^{\, \epsilon_1,\epsilon_2}_C(x;N) G^{-\epsilon_1,\epsilon_2}_C(x;N).\eeq
Combining this with (\ref{Gbound}) we obtain the useful bound
\beq  P_C^{(N)}(x)\le \frac{3 b_-}{b_+^2}\exp(-N(\Lambda^{\epsilon_1,\epsilon_2}(x)+\Lambda^{-\epsilon_1,\epsilon_2}(x))).\label{PNbound}\eeq

Now consider the ensemble $\mu'$ of  combs $C$ for which: $T_k=N_1, k=1..K-1$, $T_K=N_\ell$; at $k>K$  teeth are short, $T_k=N_1$, with probability $1-p$ or long, $T_k=N_\ell$, with probability $p$; and the $n$th tooth is short, $T_n=N_1$. Then using the representation (\ref{GCreln}) $G_C(x;n)$ can be bounded above by noting that if $T_{k+1}=N_\ell$ then $P_{C_k}<P_{\ell\sharp }$, otherwise $P_{C_k}<P_{\sharp}$. This gives
\bea G_C(x,n) &\le& \frac{3}{b_+}(1-x)^{-n/2}\left(\frac{b_+}{b_-}\right)^nP_{\ell\sharp }(x)^{n-K-k}  P_{\sharp }(x)^{k+K},\label{cuteGa}\eea
and hence
\bea\fl\left\langle G_C(x,n)\right\rangle_{\mu'}
&=&\sum_{k=0}^{n-K-1}{\scriptstyle {n-K-1}\choose{k}}\, p^{n-K-1-k}  (1-p)^{k} G_C(x,n)\nn\\
 &\le&
 \frac{3}{b_+} \left(\frac{b_+}{b_-}\right)^n\frac{P_{\sharp}(x)^{K}P_{\ell\sharp}(x)}{   (1-x)^{n/2}   }
\left((1-p)P_{\sharp}(x)+pP_{\ell\sharp }(x)\right)^{n-K-1}.\label{cuteG}\eea

\section{Results independent of the comb ensemble $\mu$}

In this section we show that in some regions of $\epsilon_{1,2}$ the behaviour  at large time is essentially independent of the comb ensemble, or else simply dependent upon $\avg{P_T(x)}$.  The leading, and where different, the leading non-analytic, behaviour of $\avg{ P_T(x)}$ as $x\to 0$ for the measures studied here is given in  table \ref{Table2}.  The results for $\mu^A$ are trivial, as are those for any measure when $\epsilon_2<0$, while the case $\mu^B$ and $\epsilon_2=0$ can be derived using the techniques in \cite{comb}.  The calculation for $\mu^B$ and $\epsilon_2 >0$ is somewhat subtle and is included in \ref{AppPT}.  

\begin{table}[h!]
\caption{\label{Table2}Leading and leading non-analytic behaviour of $1-\avg{P_T}$ in various cases.}
\begin{indented}
\item[]\begin{tabular}{@{}llll}
\br
ensemble&$\ep_2<0$&$\ep_2=0$&$\ep_2>0$\\
\mr
$\mu^A$& $Bx$&$B x^{\half}$&$B+B' x$\\
$\mu^B, a<2$&  $Bx$&  $Bx^{a/2}$ &$B (\abs{\log x}^{a-1})^{-1}$\\$\mu^B, a=2k$&  $Bx$&  $Bx+\ldots B'x^k\abs{\log x}$ &$B (\abs{\log x}^{a-1})^{-1}$\\
$\mu^B, a>2, a\ne2k$&  $Bx$&  $Bx+\ldots B'x^{a/2}$ &$B (\abs{\log x}^{a-1})^{-1}$\\
\br
\end{tabular}
\end{indented}
\end{table}

\subsection{$d_s$ when $\epsilon_2<0$ }
First we show that for any comb ensemble 
\beq d_s=\cases{0 & {if $\epsilon_1<0$ and $\epsilon_2<0$};\\
1  & {if $\epsilon_1=0$ and $\epsilon_2<0$.}}\label{res0}\eeq
By monotonicity we have that for any comb $C$
\beq P_*(x)\le P_C(x)\le P_{\sharp}(x).\eeq
Taking expectation values and using (\ref{Psharp}) and (\ref{PstarB}) it follows that for $\epsilon_2<0$
\beq\fl P(x)=\avg{P_C(x)}=\cases{1-B_1 x^\half + O(x) & {if $\epsilon_1=0$,}\\
                               \frac{1-\epsilon_2-\abs{\epsilon_1}}{b_+}-x\frac{ B_2}{\abs{\epsilon_1}} +O(x^2)& {otherwise.}}\label{res1}\eeq
Similarly 
\beq Q_*(x)\le Q_C(x)\le Q_{\sharp}(x)\eeq
and so
\beq\fl Q(x)=\avg{Q_C(x)}=\cases{\frac{B_1}{ x^\half}+O(1) & {if $\epsilon_1=0$,}\\
                               \frac{ B_2\abs{\epsilon_1}}{x} +O(1)& {if $\epsilon_1<0$,}}\label{res2}\eeq
and (\ref{res0}) follows. 

\subsection{$d_s$ when  $\epsilon_1>0$\label{nonrecurrent}}

When $\epsilon_1>0$ all combs are non-recurrent and so we must examine the derivatives of $Q(x)$. Differentiating (\ref{QPreln}) and (\ref{recur}) gives
\bea Q^{(1)}_C(x)&=&Q_C(x)^2 P^{(1)}_C(x),\label{nr1}\\
P^{(1)}_C(x)&=&\frac{-P_C(x)}{1-x}+\frac{P_C(x)^2}{(1-x)b_-}\times\nn\\
&&\qquad\left(
b_TP^{(1)}_{T_1}(x)+b_+P^{(1)}_{C_1}(x)\right).\label{Pdiff}\eea
By monotonicity (\ref{Pdiff}) can be bounded above and below by replacing $P_C$ with $P_*$ and $P_\sharp$ respectively. Taking the expectation value and using translation invariance to note that $\avg{P_C}=\avg{P_{C_1}}$
shows that, if $\avg{P^{(1)}_{T}(x)}$ diverges as $x\to 0$, then 
\beq {Q^{(1)}(x)}\sim B \avg{P^{(1)}_{T}(x)} + O(\sqrt{x} \avg{P^{(1)}_{T}(x)},1).\label{nr2}\eeq
As can be seen from table \ref{Table2}, in some cases $\avg{P_T(x)}$ is analytic, or only higher derivatives diverge. For the measures considered here it can be shown that if $\avg{P_T(x)}$ is analytic at $x=0$ then so is  $Q(x)$. If on the other hand
$\avg{P_T(x)}$ is not analytic
 but  the $k$'th derivative diverges then
\beq {Q^{(k)}(x)}=B\avg{P^{(k)}_{T}(x)}++ O(\sqrt{x} \avg{P^{(k)}_{T}(x)},1).\label{Qkdiv}\eeq
The proof is a straightforward but tedious generalization of (\ref{nr1}) and  (\ref{Pdiff}) and is relegated to \ref{AppNR}. If a derivative of $Q(x)$ diverges then $d_s$ can be read off using (\ref{dsdefn}) and (\ref{Qkdiv}). Otherwise if all finite order derivatives are finite then $p_C(t)$ decays at large $t$ faster than any power and \(d_s\) is not defined.  

\subsection{$d_k$ when $\epsilon_2<0$ or $\epsilon_1<0$}
 We  show that for any comb ensemble 
\beq 
\tilde{d_k}=0, \quad d_k=\cases{0 & {if $\epsilon_1<0$,}\\
k/2  & {if $\epsilon_1=0$ and $\epsilon_2<0$,}\\
k  & {if $\epsilon_1>0$ and $\epsilon_2<0$.}
}\label{res0a}
\eeq

It is trivial to show that 
\beq 1\le D_\ell\le \frac{B}{\abs{\ep_2}},\quad \epsilon_2<0,\eeq
and then by monotonicity we get
\beq \frac{G_*(x;n)}{1-P_*(x)}\le H(x;n)\le \frac{B}{\abs{\ep_2}} \frac{G_\sharp(x;n)}{1-P_\sharp(x)}.\eeq
Combining this with (\ref{Psharp}) and (\ref{PstarB}) yields the results for $\epsilon_2<0$.

To deal with $\epsilon_1<0$ and $\epsilon_2\ge 0$ note that monotonicity gives
\beq \avg{ \frac{D_{\abs{ T_n}}(x)}{1-P_C(x)}}\,  G_*(x;n)   \le  H(x;n)\le
\avg{ \frac{D_{\abs{ T_n}}(x)}{1-P_C(x)}} \, G_\sharp(x;n).\label{intermediate} \eeq
Using the lower bound and    (\ref{GCreln}), (\ref{unitarity}) and (\ref{PstarA}) we get after summing over $n$
 \beq \avg{ \frac{D_{\abs{ T_n}}(x)}{1-P_C(x)}}\, \frac{3}{b_+} \frac{b_-}{\sqrt{4\ep_2+\ep_1^2}-\ep_1-2\ep_2}   \le \sum_{n=0}^\infty H(x;n)\le \frac{2}{x}.\eeq
Inserting this into the upper bound of (\ref{intermediate}) gives
\beq H(x;n) \le  \frac{2}{x}\frac{b_+}{3} \frac{\sqrt{4\ep_2+\ep_1^2}-\ep_1-2\ep_2} {b_-}\,  G_\sharp(x;n).\eeq
It is a trivial consequence of (\ref{unitarity}) that
\beq \sum_{n=0}^\infty n^k H(x;n) >\frac{c}{x},\qquad k>0,\eeq
and the results then follow by using (\ref{Gbound}).

\section{The spectral dimension  when $\epsilon_2\ge 0$ and $\epsilon_1\le0$}
Here and in some of the sections to follow we will need to sum over the location of the first long tooth to determine the spectral dimension.  Most generally we call a tooth long when it has length \(\ge \ell\) and short when it has length \(<\ell \).  Consider combs for which the first $L-1$ teeth are short but the $L$th tooth is long; the probability for this is $p(1-p)^{L-1}$, where \(p\) is the probability of a tooth being long.  Denoting by \(\ell L\) a comb having the first long tooth at vertex \(L\) gives 
\beq Q(x) = \sum_{L=1}^\infty \avg{Q_{\ell L}(x)} p(1-p)^{L-1}.\label{Qsum}\eeq
$Q_{\ell L}(x)$ is bounded above by the comb in which all teeth at $n \ge L+1$ are short, and below by the comb in which all teeth at $n\ge L+1$ are infinite,
\beq Q_{\{T_{n<L}=N_{\ell'},\ell' \le \ell; T_{n\ge L}=N_\infty\}}(x)<  Q_{\ell L}(x) < Q_{\{T_{n \ne L}=N_1;T_L=N_\ell\}}(x).\label{submaster}\eeq

\subsection{$\mu^A$ -- Infinite teeth at random locations\label{ITRL}}

\subsubsection{$\epsilon_2= 0$, $\epsilon_1< 0$\label{ITRL1}}
We first show that  the exponent $\beta=\half$ -- so it is unchanged from the comb $*$. This result follows from the inequalities
\beq 1-\frac{pBx^\half}{\abs{\epsilon_1}}+O(x)\le P(x)\le 1-pB'x^\half +O(x).\eeq
The lower bound is obtained by applying Jensen's inequality to (\ref{recur}).
To get the upper bound we average over the first tooth and  then by monotonicity we obtain
\beq P(x)\le p P_{\ell\sharp}(x) + (1-p) P_{\sharp}(x),\label{Ptop}\eeq
with $\ell=\infty$ and using (\ref{recur}) and (\ref{Psharp}) gives the bound required.

The spectral dimension is given by
\beq d_s = \cases{1 & {if $p\ge {2\abs{\epsilon_1}}({1+\abs{\epsilon_1}})^{-1}$,}\\
                  \frac{\log(1-p)}{\log\left(\frac{1-\abs{\epsilon_1}}{1+\abs{\epsilon_1}}\right)}  & {otherwise.}}\eeq
This result follows from estimating the sum in (\ref{Qsum}) using the bounds in (\ref{submaster}) with \(\ell=\infty\) and short teeth being \(N_1\).
$P_C(x) $ for these bounding combs is computed in \ref{AppL} and using (\ref{PLupper}) we get upper and lower bounds on \(Q_{\infty L}(x) \) of the form
\beq \frac {1}{Bx+B'x^{\half}\left(\frac{1-\abs{\epsilon_1}}{1+\abs{\epsilon_1}}\right)^L}.\eeq

\subsubsection{$\epsilon_2> 0$, $\epsilon_1< 0$\label{ITRL2}}

The probability that  $C$ is non-recurrent is at least $p$, the probability that $T_1=N_\infty$, and hence 
\beq P(0)< 1.\eeq
In fact it follows from the lemma of \ref{AppNR} that $P^{(k)}(x)$ is finite for all finite $k$ so the exponent $\beta$ is undefined.

The spectral dimension is given by
\beq d_s=\frac{2\log(1-p)}{\log\left(\frac{1-\abs{\epsilon_1}-\ep_2}{1+\abs{\epsilon_1}-\ep_2}\right)}.\label{s33}\eeq
To show this we start by estimating $Q(x)$ in exactly the same way as in  \ref{ITRL1}  except that the behaviour of the limiting combs is now given by (\ref{PLupper2}) so that there are upper and lower bounds on \(Q_{\infty L}(x) \) of the form
\beq
\frac {1}{Bx+B'\left(\frac{1-\abs{\epsilon_1}-\ep_2}{1+\abs{\epsilon_1}-\ep_2}\right)^L}.\label{Qres2}
\eeq
When $p\le1-b_+/b_-$ this sum diverges at $x=0$
 and it is then straightforward to obtain (\ref{s33}). For larger $p$ the sum is convergent at $x=0$ so we next examine $Q^{(1)}(x)=\avg{Q_C^2P_C^{(1)}}$. Note that $-P_C^{(1)}\ge \third b_-$; then letting $Z$ be a very large integer and using H\"older's inequality 
\beq b_-\avg{Q_C(x)^2}\le- Q^{(1)}(x) \le \avg{Q_C(x)^{2+1/Z}}^{\frac{2Z}{2Z+1}}
\avg{-P_C^{(1)}(x)^{2Z+1}}^{\frac{1}{2Z+1}}.\eeq
By the lemma of \ref{AppNR} the second factor in the upper bound is finite as $x\to0$ so we need an estimate of $\avg{Q_C^2}$. This is provided by (\ref{Qres2}) modified by squaring the denominator;   when $p\le1-(b_+/b_-)^2$ this sum diverges at $x=0$ and once again we  obtain (\ref{s33}).  For still larger $p$ both $Q$ and $Q^{(1)}$ are finite at $x=0$ and we examine the second  and higher derivatives. This uses (\ref{Qnderiv}), $(-1)^kP_C^{(k)}\ge  b_-^kb_+^{k-1} / 3^{2k-1} $, H\"older's inequality and the lemma; the term with the highest power of $Q_C$ dominates  and the result is always (\ref{s33}). \footnote{Strictly speaking when \(1-(b_+ / b_-)^k <p \le 1-(b_+/ b_-)^{k+1/Z} \) the upper bounds diverge so our proof does not work for these arbitrarily small intervals.}

\subsubsection{$\epsilon_2> 0$, $\epsilon_1= 0$\label{ITRL3}}
By the same argument as in \ref{ITRL2} we find $P(0)<1$, so $\beta$ is again undefined.  An upper bound on $Q(x)$ may be obtained as in \ref{ITRL1} using (\ref{PLupper4}) to get
\bea Q_{\infty L}(x)&\le&(L+(1-\ep_2)/4\ep_2)\eea
which means the upper bound of (\ref{Qsum}) is finite. A proof that all derivatives of $Q(x)$ are finite is given in \ref{last}, so $p_C(t)$ decays faster than any power at large $t$.

\subsection{$\mu^B$ -- Teeth of random length}
In this subsection we are concerned with random combs that have a distribution of tooth lengths.   The general strategy for determining quantities of interest is to identify teeth that are long enough to affect the critical behaviour of the biased random walk and consider the probability with which they occur.  It will be useful to define the function
\beq \lambda (\delta,\eta,\zeta) = \lfloor \frac{\delta\abs{\log x}^\eta-\zeta(a-1)\log\abs{\log x}}{\log Y} \rfloor, \eeq
which will be used to denote a tooth length, and the function
\beq
p_{_>}(\ell)=\sum_{k=\ell}^{\infty}\frac{C_a}{k^a}=\frac{C_a}{(a-1)\ell^{a-1}}\left(1+O(\ell^{-1})\right),
\eeq   
which is the probability that a tooth has length greater than \(\ell-1\).  

\subsubsection{$\epsilon_2= 0$,  $\epsilon_1< 0$}
We first show that 
\beq \beta = \cases{\frac{a}{2} & {if $a<2$,}\\
                      1 & {otherwise.}}\eeq 
The proof follows the lines described in section \ref{ITRL1} with a slight modification for the upper bound on $P(x)$. 
Note that, from (\ref{Pell}), teeth of length $\ell > \lfloor x^{-\half} \rfloor $ have $P_T(x)\le 1-B x^\half$.
We then proceed as in (\ref{Ptop}) but with $\ell=\lfloor x^{-\half} \rfloor+1$.  

The exponent $\beta$ is non-trivial if $a<2$ but, as we now show, $d_s=0$ for all $a>1$ so mean field theory does not apply when \(a<2\).  
This result follows from the inequalities
\beq \frac{B'}{x\abs{\log x}^{\frac{1}{a-1}}} \le Q(x)\le \frac{B}{x}.\eeq
The upper bound is a consequence of $Q(x)<Q_{\sharp}(x)$. To obtain the lower bound consider the combs for which at least the first $N$ teeth are all shorter than $\ell_0$. 
Then
using monotonicity and (\ref{PNbound})
\beq Q(x)\ge\frac{(1-p_{_>}(\ell_0))^N}{1-P_{\flat\ell_0}(x) +O(\exp(-N(\Lambda^{\epsilon_1,\epsilon_2}(x)+\Lambda^{-\epsilon_1,\epsilon_2}(x))))}.\label{Qlower}\eeq
Setting $\ell_0=\lambda(1,(a-1)^{-1},0)$, $N=\lfloor 2(\Lambda^{\epsilon_1,\epsilon_2}+\Lambda^{-\epsilon_1,\epsilon_2})^{-1}\abs{\log x} \rfloor +1$ and using (\ref{PflatA}) the result follows for small enough $x$.  

\subsubsection{$\epsilon_2> 0$, $\epsilon_1 <0$ \label{RLUL}}

The exponent $\beta=0$ but there are computable logarithmic corrections and we find that
\beq 1-\frac{B}{\abs{\log x}^{a-1}}\le P(x)\le  1-\frac{B'}{\abs{\log x}^{a-1}}.\eeq
The lower bound follows from applying Jensen's inequality to (\ref{recur}).
For the upper bound note that teeth of length
$\ell > \lambda(1,1,0)$ have $P_T<B$.  Again proceed as in (\ref{Ptop}) with $\ell=\lambda(1,1,0) +1$.

The spectral dimension is $d_s=0$ showing again that mean field theory does not apply.  
This follows from the inequalities
\beq \frac{B'\exp\left(-B''\abs{\log x}^{1/a}\right)}{x}\le Q\le \frac{B}{x},\eeq
for small enough $x$.  The upper bound is a consequence of $Q(x)<Q_{\sharp}(x)$ and the lower bound follows from  (\ref{Qlower}) by setting $\ell_0=\lambda(1,1/a,0)$,
$N= \lfloor 2(\Lambda^{\epsilon_1,\epsilon_2}+\Lambda^{-\epsilon_1,\epsilon_2})^{-1}\abs{\log x} \rfloor +1$ and using (\ref{PflatB}). 

\subsubsection{$\epsilon_2> 0$, $\epsilon_1 =0$ \label{RLUM}}
The exponent $\beta=0$, but there are logarithmic corrections which follow from the inequalities 
\beq 1-\frac{B}{\abs{\log x}^{(a-1)/2}}\le P(x)\le  1-\frac{B'}{\abs{\log x}^{(a-1)/2}}.\eeq
The lower bound comes from applying Jensen's inequality to the recurrence relation \eqref{recur}.  The upper bound is obtained by requiring unitarity of the heat kernel and its proof is relegated to \ref{AppPc}.  

The spectral dimension and logarithmic exponent are given by
\bea d_s = 2,\nn\\ \tilde\alpha=a-1,\eea
which shows that mean field theory does not apply.  This result follows from
\beq B'\,\abs{\log x}^{a-1} < Q(x) < B\,\abs{\log x}^{a-1}\label{Qres10}\eeq
for small enough $x$ which is obtained by a modified version of the argument in \ref{ITRL1}. First let
$\ell_0=\lambda(1,1,\zeta)$,
so that
\beq P_{\ell_0}(x)=1-\frac{B}{\abs{\log x}^{\zeta(a-1)}} +O\left(\frac{1}{\abs{\log x}^{2\zeta(a-1)}}\right).\eeq
To obtain (\ref{Qres10}) we use (\ref{Qsum}) and (\ref{submaster}) with \(p=p_>(\ell_0)\), \(\ell=\ell_0\) and for the lower bound set \(T_{n < L}=N_{\ell_0}\).
Then using the bounds in (\ref{PLupper3}) with $\zeta=1$ and (\ref{PLlower3}) with $\zeta=2$ and estimating the sums gives the result.

\section{Heat Kernel when $\epsilon_1 \ge0$,  $\epsilon_2 \ge0$}

These calculations require $\avg{D_\ell}$ in the various cases which are tabulated in table \ref{Table1} for convenience.

\begin{table}
\caption{\label{Table1}$\avg{D_\ell}$ in various cases.}
\begin{indented}
\item[]\begin{tabular}{@{}llll}
\br
ensemble&$\ep_2<0$&$\ep_2=0$&$\ep_2>0$\\
\mr
$\mu^A$& $B+O(x)$&$B x^{-\half}+O(1)$&$B x^{-1}+O(1)$\\
$\mu^B, a\ge2$&  $B+O(x)$&  $B+O(x)$ &$B (x\abs{\log x}^{a-1})^{-1}+O(1)$\\
$\mu^B, a<2$&  $B+O(x)$&  $Bx^{a/2-1}+O(1)$ &$B (x\abs{\log x}^{a-1})^{-1}+O(1)$\\
\br
\end{tabular}
\end{indented}
\end{table}

\subsection{$\mu^A$ -- Infinite teeth at random locations\label{ITRLHK}}

We  show that 
\beq d_k=\cases{0 & {if $\epsilon_2>0$ and $\epsilon_1 \ge0$,}\\
k/2  & {if $\epsilon_2=0$ and $\epsilon_1>0$.}
}\label{res10a}\eeq
These results follow from (\ref{H1}), (\ref{H2}) and (\ref{H3}) below.

Noting that for  $\ep_1>0$ all combs have $1-B_{-}^{-1}-<P_C(x)<1-B_{+}^{-1}$  and using monotonicity gives
\bea B_- \avg{ D_{\abs{ T_n}}(x)}\,  G_*(x;n)  \le  H(x;n)&\le& 
\avg{ D_{\abs{ T_n}}(x)} \, \avg{\frac{G_{C'}(x;n)}{1-P_{C'}(x)}},\label{intermediate0}\\
&\le& B_+
\avg{ D_{\abs{ T_n}}(x)} \, \avg{G_{C'}(x;n)},\label{intermediate1} \eea
where $C'$ is constructed from $C$ by forcing $T_n=N_1$. If $\ep_2>0$ then using 
(\ref{cuteG}) with $K=0, \ell=\infty$ gives the upper bound 
\beq H(x;n) < \frac{B}{x}\exp(-B'n).\label{H1}\eeq
 If $\ep_2=0$ then exactly the same calculation gives
\beq H(x;n)  < \frac{B}{x^{\half}}\exp(-B'x^{\half}n)\label{H2} \eeq
and evaluating the left hand side of (\ref{intermediate0}) gives a lower bound of the same form. If $\ep_2>0$ and $\ep_1=0$  it is necessary to sum over the location of the first infinite tooth. Using (\ref{PLupper4}), (\ref{cuteG}) and introducing $C'$ as in (\ref{intermediate0}) gives
\beq H(x;n) < \frac{B}{x}\exp(-B'n).\label{H3}\eeq

\subsection{$\mu^B$ -- Teeth of random length\label{RLHK}}

We  show that 
\beq d_k=\cases{0 & {if $\epsilon_2>0$ and $\epsilon_1 \ge 0$;}\\
ka/2  & {if $\epsilon_2=0$, $\epsilon_1>0$ and $a<2$;}\\
k  & {if $\epsilon_2=0$, $\epsilon_1>0$ and $a\ge 2$.}\\
}\label{res10b}\eeq
These results follow from 
\beq\fl H(x;n) < \cases{\frac{B}{x\abs{\log x}^{a-1}}\exp(-B'n/\abs{\log x}^{a-1})  & {if $\epsilon_2>0$ and $\epsilon_1 \ge 0$;}\\
\frac{B}{x^{1-a/2}}\exp(-B'n x^{a/2})  & {if $\epsilon_2=0$, $\epsilon_1>0$ and $a<2$;}\\
B\exp(-nB'x)  & {if $\epsilon_2=0$, $\epsilon_1>0$ and $a\ge 2$,}\\
}\label{res10b1}\eeq
when $x$ is small enough and lower bounds of the same form.  

The upper bounds are obtained by proceeding as in subsection \ref{ITRLHK}: for $\ep_1>0$ and $\ep_2>0$ setting  $\ell=\lambda(1,1,0)+1$ and for $\ep_1>0$ and $\ep_2=0$ setting $\ell=\lfloor x^{-\half} \rfloor +1$.  For $\ep_1=0$ and $\ep_2>0$ we start with the upper bound of \eqref{intermediate0};
let \(\ell_1=\lambda(1,1,2)\), \(p_1=p_{_>}(\ell_1)\) and \(\ell_2=\lambda(2,1,0)\), \(p_2=p_{_>}(\ell_2)\).  The latter shall be called long teeth and we denote by \((\ell_2K\sharp)\) the comb with a single long tooth at vertex \(K\).  We now sum over the location of the first long tooth using \eqref{GCreln}, \eqref{cuteG} and \eqref{PLupper3} and taking account of the fact that the first long tooth may be before or after the \(n\)th tooth
\fl
\bea
H(x;n) & \le& \avg{ D_{\abs{ T_n}}(x)} \sum_{K=1}^{n-1} \frac{p_2(1-p_1)^{K-1}}{1-P_{(\ell_2K\sharp)}(0)} \prod_{m=0}^{K-1}P_{(\ell_2K\sharp)_{m}} \nn \\
& & \times \left((1-p_1)P_\sharp+(p_1-p_2)P_{\ell_1 \sharp}+p_2P_{\ell_2 \sharp} \right)^{n-K-1}   \nn \\
& & + \avg{\theta(\ell_2-\abs{T_n}) D_{\abs{ T_n}}(x)} \sum_{K=n}^{\infty} \frac{p_2(1-p_1)^{K-1}}{1-P_{(\ell_2K\sharp)}(0)} \prod_{m=0}^{n-1}P_{(\ell_2K\sharp)_{m}}.
\eea
In the first sum we use the value given in Table \ref{Table1} for $\avg{ D_{\abs{ T_n}}(x)}$.  In the second sum $\avg{\theta(\ell_2-\abs{T_n}) D_{\abs{ T_n}}(x)}= B (x\abs{\log x}^{2(a-1)})^{-1}+O(1)$ for \(\abs{T_n} < \ell_2\) and the result follows.  

To obtain the lower bounds when $\epsilon_1>0$ we note that 
\bea H(x;n)&\ge& B_- \avg{D_{\abs{T_n}}(x)G_C(x;n)}\nn\\
&=&B_- \avg{D_{\abs{T_n}}(x)}\avgp{G_C(x;n)}\eea
where the measure $\mubar$ is defined by
\bea \label{muBar} \mubar_\ell&=&\mu_\ell,\qquad{\rm for~teeth}\; T_k,\, k\ne n,\nn\\
\mubar_\ell&=&\frac{\mu_\ell D_\ell}{\avg{D_\ell}},\qquad{\rm for~tooth}\; T_n.\eea
Using the decomposition (\ref{GCreln}) and Jensen's inequality
\beq \avgp{G_C(x;n)}\ge \frac{3(1-x)^{n/2}}{b_+}\exp(-S_n),\eeq
where
\beq S_n=\sum_{k=0}^{n-1}\avgp{\frac{b_-}{b_+}-P_{C_{k+1}}(x)}
+\frac{b_T}{b_+}\avgp{1-P_{T_{k+1}}(x)}.\eeq
Now  applying Jensen's inequality  with the measure $\mubar$ to (\ref{recur}) shows that the lower bounds satisfy a recursion formula of exactly the same form as discussed in \ref{AppL}. So from  (\ref{Ldecay}) we find that 
\bea S_n&\le&n\left(\frac{b_-}{b_+}-P(x)+\frac{b_T}{b_+}\avg{1-P_{T}(x)}\right)\nn\\&&\quad -\frac{b_T}{b_+}(\avgp{P_{T}(x)}-\avg{P_{T}(x)})+\nn\\&&-\quad\sum_{k=1}^{n-1}\frac{P(x)(1-A(x))}{A(x)^{k-1}(\bar P(x)-A(x)P(x))/(\bar P(x)-P(x))-1},\label{crucial}\eea
where
\bea P(x)&=&\frac{(1-x)b_-}{3-b_T\avg{P_T(x)}-b_+P(x)}\;,\nn\\
\bar P(x)&=&\frac{(1-x)b_-}{3-b_T\avgp{P_T(x)}-b_+P(x)}\;,\nn\\ A(x)&=&\frac{(1-x)b_-}{P(x)^2\,b_+}.\label{crucialA}\eea
For $\ep_1>0$ it is straightforward to check that $A(x)>c>1$  and that the sum in (\ref{crucial}) is bounded above by an $n$ independent constant. Lower bounds of the form of (\ref{res10b1}) then follow by inserting the appropriate $\avg{P_T}$ in 
(\ref{crucialA}) and (\ref{crucial}).

When $\ep_1=0$
\beq H(x;n)\ge 
\avg{ D_{\abs{ T_n}}(x)} \, \avg{\frac{G_{C'}(x;n)}{1-P_{C'}(x)}},\label{intermediate2}\eeq
where $C'$ is constructed from $C$ by setting $T_{k\ge n}=N_\infty$. 
Choosing $\ell_0=\lambda(1,1,2)$ and using (\ref{GCreln}) and (\ref{PLlower3}) gives
\bea
\lefteqn{\fl H(x;n)\ge \avg{ D_{\abs{ T_n}}(x)} (1-p_>(\ell_0))^{n-1} 3 (1-x)^{-n/2}P_*(x)^2 \frac{\prod_{k=3}^{n-1}\left(P_{\flat\ell_0}(x)-\frac{1}{k-1}\right)}  {1-P_{\flat\ell_0}(x)+\frac{1}{n-1}}} \nn \\
\lefteqn{\fl \phantom{H(x;n)}  >  \avg{ D_{\abs{ T_n}}(x)} (1-p_>(\ell_0))^{n-1} 3 (1-x)^{-n/2}P_*(x)^2} \nn \\
\lefteqn{\fl \phantom{H(x;n) > }\times \frac{1}{n-2} \cdot \frac{1}{1-P_{\flat\ell_0}(x)+\frac{1}{n-1}} \exp{\left\{\frac{(n-3)2(1-P_{\flat\ell_0}(x))}{2P_{\flat\ell_0}(x)-1}\right\}},}
\eea
for $n \ge 4$ which gives the result.



\section{Results and discussion}
\begin{figure}[h!] 
\begin{center}
\centerline{\hbox{\scalebox{1}{\includegraphics{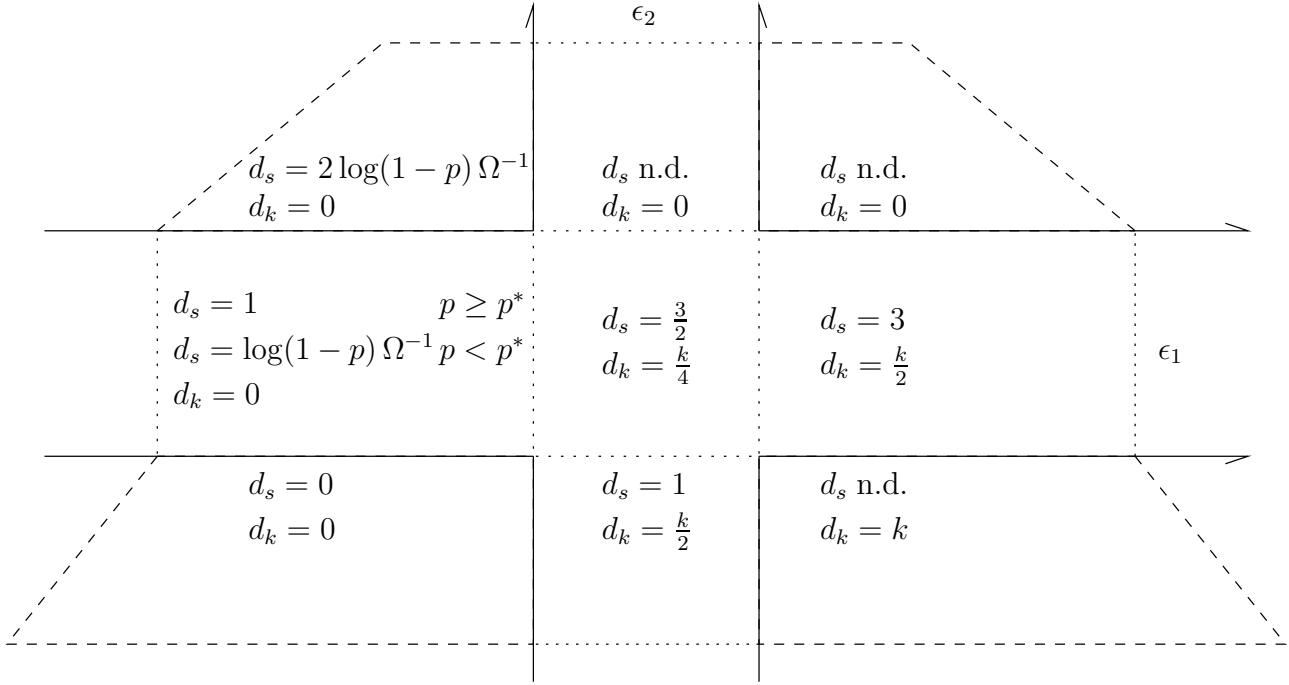}}}}
\caption{Results for $\mu^A$ where $\Omega=\log \left(\frac{1-\abs{\ep_1}-\ep_2}{1+\abs{\ep_1}-\ep_2} \right)$ and $p^*=2\abs{\ep_1}(1+\abs{\ep_1}-\ep_2)^{-1}$.  The logarithmic exponents $\tilde{\alpha}$ and $\tilde{d_k}$ are always zero for $\mu^A$.  }
\label{fig3}
\end{center}
\end{figure}

Figure \ref{fig3} outlines the results that we have computed for \(\mu^A\).  These are new and show that the most interesting regime is actually when the bias along the spine is \emph{towards} the origin, a circumstance which has not been studied much in the literature.  When \(\epsilon_1 \ge 0\) and \(\epsilon_2 >0\) the walker disappears rapidly, never to return, and \(p(t)\) decays faster than any power.  When \(\epsilon_1 <0\) the bias along the spine is keeping the walker close to the origin \emph{but} if there are any infinite teeth present the walker can spend a lot of time in the teeth; the conflict between these effects leads to a non-trivial \(d_s\).  The fact that \(d_1=0\) whenever \(\epsilon_2>0\) shows that the walker never gets far down the spine; if she disappears then it is up a tooth that she is lost.  The Hausdorff dimension for $\mu^A$ is $d_H=2$, regardless of bias and so we have here several examples of violation of the bound  $2d_H/(1+d_H) \le d_s \le d_H$, which applies for unbiased diffusion \cite{Point}.  

\begin{figure}[h!] 
\begin{center}
\centerline{\hbox{\scalebox{1}{\includegraphics{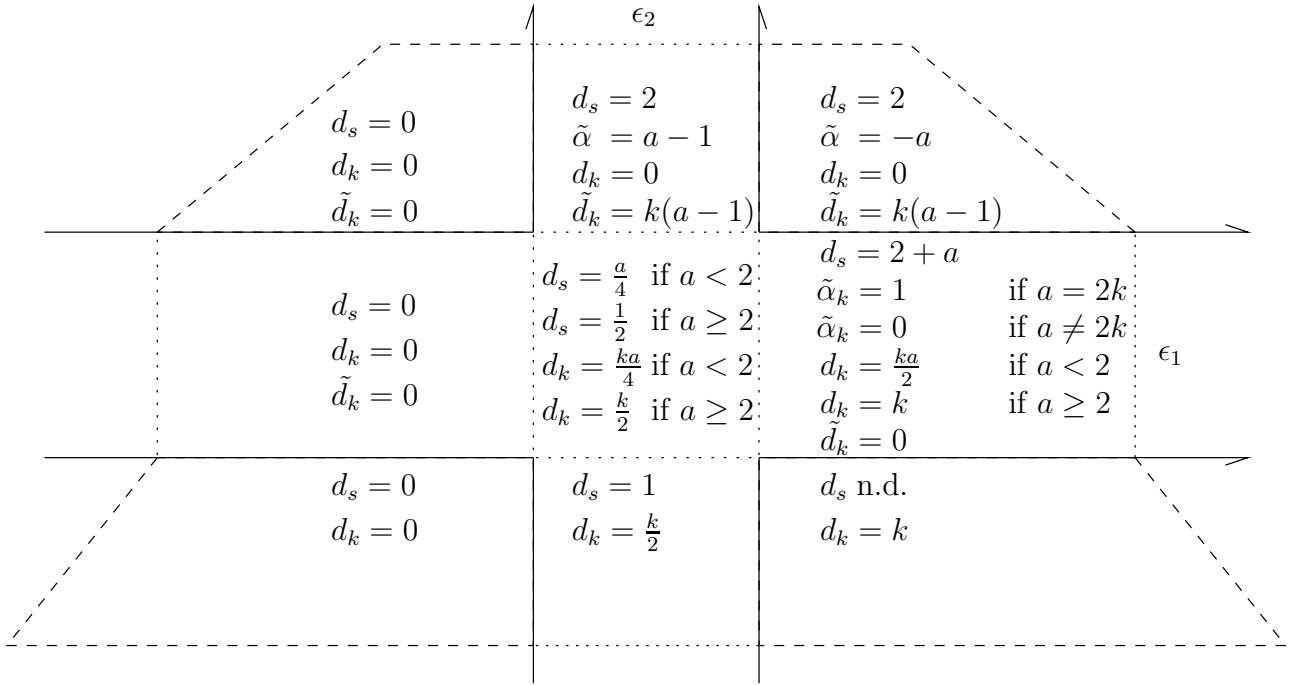}}}}
\caption{Results for $\mu^B$.  When $\ep_2 <0 $ the logarithmic exponents $\tilde{\alpha}$ and $\tilde{d_k}$ are always zero.}
\label{fig4}
\end{center}
\end{figure}

Figure \ref{fig4} shows our results for \(\mu^B\) as well as the results for the unbiased case studied in \cite{comb}.  This length distribution has been studied quite extensively in the literature but usually under the assumption that \(\epsilon_2 \ge 0\).  As can be seen the interesting behaviour displayed by \(\mu^A\) when \(\epsilon_1 <0 \) does not occur here -- essentially because very long teeth are not common enough.  We believe that with more work $\tilde{\alpha}$ when $\ep_1 <0$, $\ep_2 \ge 0$ can be found using our methods, but as this will not give further physical insight we leave the calculations to elsewhere \cite{TanyaThesis}.
The case \(\epsilon_1>0\), \(\epsilon_2 >0\) (often called topological bias) was originally studied using mean field theory, which gave the mean square displacement
\begin{equation} \label{r1}
\langle n^2(t) \rangle \sim (\log {t})^{2(a-1)},
\end{equation}
and this is in fact correct since the walker spends much of the time in the teeth.  However the claim in \cite{three} that \eqref{r1} holds for \(\epsilon_2 >0\) regardless of \(\epsilon_1\) is false.  The mean field method gives the correct result when $\ep_1=0$ only because the walk on the spine is ignored, which amounts to using $P_T(x)$ for $P_C(x)$ in \eqref{heatk} and naively applying Jensen's inequality.  The case \(\epsilon_1>0\), \(\epsilon_2=0\) was studied by Pottier \cite{Pot2} who computed the leading contribution exactly, but without complete control over the sub-leading terms; she also calculated the leading behaviour \(\langle n^2 \rangle - \langle n \rangle ^2\) which we have not.  Of course our results for \(d_s\) and \(d_1\) agree with hers.   The Hausdorff dimension for $\mu^B$ is $d_H=3-a$ when $a<2$ and $d_H=1$ when $a \ge 2$ and so again we see that, as expected, a biasing field intensifies the difference between the purely geometric definition of dimension and that which is related to particle propagation.  

The results for \(\epsilon_2 <0\) are intuitively obvious and, as we have proved, apply for any model with identically and independently distributed tooth lengths.  The walker never gets far into the tooth and therefore combs have long time behaviour characteristic of the spine alone.  

This paper has given a comprehensive treatment of biased random walks on combs using rigorous techniques -- namely recursion relations for generating functions combined with unitarity and monotonicity arguments.  It serves to put in context many previous results as well as present new ones.  In the unbiased case \cite{comb} and in some bias regimes mean field theory is sufficient to compute the leading order behaviour because the walker either does not reach the ends of the longest teeth or does not travel far enough down the spine for variations from average to be important.  But, as is illustrated in many examples here, a full treatment is needed when such fluctuations cannot be ignored.  Finally, while the results are of interest in themselves, an important point of the paper was to demonstrate that rigorous analytic methods can be used to treat biased diffusion on random geometric structures and it is to be hoped that these tools can be extended to higher dimensional problems.  

\psfrag{ds0}{$d_s=0$}
\psfrag{dk0}{$d_k=0$}
\psfrag{ds1}{$d_s=1$}
\psfrag{ds3}{$d_s=3$}
\psfrag{dkk}{$d_k=k$}
\psfrag{dkk/2}{$d_k=\frac{k}{2}$}
\psfrag{ds3/2}{$d_s=\frac{3}{2}$}
\psfrag{ds1/2}{$d_s=\frac{1}{2}$}
\psfrag{dkk/4}{$d_k=\frac{k}{4}$}
\psfrag{dsnd}{$d_s \; \textrm{n.d.}$}
\psfrag{ds2om}{$d_s=2 \log (1-p) \, \Omega^{-1}$}
\psfrag{ds1om}{$d_s= \log (1-p) \, \Omega^{-1}$}
\psfrag{pl}{$p<p^{*}$}
\psfrag{pge}{$p \ge p^{*}$}

\psfrag{ds2}{$d_s=2$}
\psfrag{dsa/4}{$d_s=\frac{a}{4}$}
\psfrag{ds(2+a)}{$d_s=2+a$}
\psfrag{altle0}{$\tilde{\alpha} \; \le 0$}
\psfrag{altk1}{$\tilde{\alpha}_k =1$}
\psfrag{altk0}{$\tilde{\alpha}_k =0$}
\psfrag{alta-1}{$\tilde{\alpha} \; =a-1$}
\psfrag{alt-a}{$\tilde{\alpha} \; =-a$}
\psfrag{dkt0}{$\tilde{d_k}=0$}
\psfrag{dktk(a-1)}{$\tilde{d_k}=k(a-1)$}
\psfrag{altb}{$\frac{-1}{a-1} \le \tilde{\alpha} \le 0$}
\psfrag{dkka/2}{$d_k=\frac{ka}{2}$}
\psfrag{dkka/4}{$d_k=\frac{ka}{4}$}
\psfrag{al2}{if $a<2$}
\psfrag{age2}{if $a \ge 2$}
\psfrag{a2k}{if $a = 2k$}
\psfrag{an2k}{if $a \ne 2k$}

\ack
We would like to thank Bergfinnur Durhuus and Thordur Jonsson for valuable discussions.  This work is supported in part by Marie Curie grant MRTN-CT-2004-005616 and by UK PPARC grant PP/D00036X/1.  T.E. would like to acknowledge an ORS award and a Julia Mann Graduate Scholarship from St Hilda's College, Oxford.  

\appendix

\section{Calculation of $\avgB {P_T(x)}$ for $\epsilon_2>0$\label{AppPT}}
First we rewrite (\ref{Pell}) as
\beq
P_\ell(x) = P_\infty(x)Y - (Y-1)P_\infty(x)X^{-1}\frac{1}{X^{-1}+Y^{-\ell}},
\eeq
so that 
\beq
\avgB {P_T(x)} = P_\infty(x)Y - (Y-1)P_\infty(x)X^{-1} \avgB {\frac{1}{X^{-1}+Y^{-\ell}}}
\eeq
with 
\beq
\avgB {\frac{1}{X^{-1}+Y^{-\ell}}}=\sum_{\ell=1}^\infty \frac{C_a \ell^{-a} }{X^{-1}+Y^{-\ell}} \equiv S.
\eeq
Since for $\epsilon_2 >0$, $Y >1$ we let $\log Y=\rho$ and write
\beq \label{Ssum}
S = \sum_{\ell=1}^{\lfloor\sigma\frac{\abs{\log x}}{\rho} \rfloor} \frac{C_a \ell^{-a} }{X^{-1}+e^{-\rho\ell}} + \sum_{\lfloor \sigma\frac{\abs{\log x}}{\rho} \rfloor +1}^{\infty} \frac{C_a \ell^{-a} }{X^{-1}+e^{-\rho\ell}},
\eeq
where \(\sigma\) is an arbitrary constant \(<1\).  This is bounded above by taking \(\ell\) in the exponential to be its value at the top of each sum to give
\bea
S &\le& \frac{C_a }{X^{-1}+x^{\sigma}} \sum_{\ell=1}^{\lfloor\sigma\frac{\abs{\log x}}{\rho} \rfloor}  \ell^{-a}  + \frac{C_a }{X^{-1}} \sum_{\lfloor \sigma\frac{\abs{\log x}}{\rho} \rfloor +1}^{\infty}  \ell^{-a} \nn \\
S &\le& \frac{1}{x^{\sigma}}+\frac{c_0 X}{\abs{\log x}^{a-1}}.
\eea 
Noting that as \(x \to 0\), \(X^{-1} \to B x\) we get a lower bound on $\avgB {P_\ell(x)}$ of
\beq
\avgB {P_T(x)} \ge 1-\frac{B_1}{\abs{\log x}^{a-1}},
\eeq
for small enough $x$.  An equivalent upper bound is calculated in the same manner by ignoring the first term in (\ref{Ssum}) and setting \(\sigma=1\), which leads to the result quoted in table \ref{Table2}.  A similar procedure leads to bounds of the form $B/ x \abs{\log x}^a$ on $\avgB {P_T^{(1)}(x)}$, which we also need, at small enough $x$.

\section{Proof of results for non-recurrent regime\label{AppNR}}

First we define a structure of ordered lists of ordered integers. Let $S$ denote an ordered  list of $h_S$  integers
\beq S=\cases{[n_1,n_2,\ldots n_{h_S}], \quad n_1\ge n_2\ge\ldots\ge n_{h_S}\ge 1,\quad h_S\ge1,\\[\,],\quad h_S=0.}\eeq
Define 
\beq\abs S =\cases{\sum_{i=1}^{h_S}n_i,&$h_S\ge1$,\\0,&$h_S=0$,}\eeq
and let ${\cal S}^N$ denote the set of all distinct lists $S$ with $\abs S =N$. Within ${\cal S}^N$ the lists $S$ and $S'$ are ordered by letting $j=\min(i: n_i\ne n_i')$ and then setting $S>S'$ if $n_j>n_j'$. Finally if $S\in{\cal S}^N$ and $S'\in{\cal S}^{N'}$ with $N>N'$ then $S>S'$. It is convenient to denote by $S+1$ the lowest list above $S$, and by $S\cup S'$ the list obtained by concatenating $S$ and $S'$ and then ordering as above.

Now define
\beq H(S;f(x))=(-1)^{\abs S}\prod_{i=1}^{h_S}f^{(n_i)}(x),\eeq
and for the empty list $H([\,];f(x))=1$.
We need the following lemma, which is proved in \ref{Lproof}:\\
\noindent{\bf Lemma}
{\it \begin{enumerate}
\item If $\avg{H(S;P_T(x))}$ is finite as $x\to0$ for all $S\le\bar S$ then $\avg{H(S;P_C(x))}$ is finite as $x\to0$ for all $S\le\bar S$ and $\epsilon_1 \ne 0$.
\item If the conditions of part (i) apply and, as $x\to 0$, $\avg{H(\bar S+1,P_T(x))}$ diverges as $x^{-\gamma}, \gamma>0$, then $\avg{H(\bar S+1,P_C(x))}$ also diverges  as $x^{-\gamma}$.
\end{enumerate} }

Differentiating (\ref{QPreln}) $k$ times gives
\beq  Q_C^{(k)}(x)= \frac{ P_C^{(k)}(x)}{(1- P_C(x))^2}
+(-1)^k\sum_{S\in{\cal S}^k/[k]}\frac{C(S) H(S;P_C(x))}{(1- P_C(x))^{h_S+1}}\label{Qnderiv}\eeq
where $C(S)$ is a combinatorial coefficient. 
It is straightforward to check for any $S$ that  $\avgA{H(S;P_T(x))}$ is analytic for $\ep_2\ne 0$, and that $\avgB{H(S;P_T(x))}$ is analytic when  $\ep_2< 0$. When $\ep_2=0$
\beq H(S;P_\ell(x))\vert_{x=0}=c_S\ell^{2\abs{S}-h_S}\left(1+O(l^{-2})\right)\eeq
from which $\avgB{ H(S;P_\ell(x))}$ is divergent for $S=[\lceil{a/2}\rceil]$, and with smaller degree for $[\lceil{a/2}\rceil-1,1]$ if $2k<a\le2k+1,\,k\in{\mathbb Z}$, but always convergent for any inferior $S$.
The results given in section \ref{nonrecurrent} then
follow from noting that $P_*(x)<P_C(x)<P_\sharp(0)<1$ and using the lemma.

\subsection{Proof of lemma\label{Lproof}}
To prove the lemma 
 note that
\beq H(S;f+g)=\sum_{S'\cup S''=S}H(S';f)H(S'';g)\eeq
and
 differentiate (\ref{recur}) $n$ times to get 
\beq (-1)^n P_C^{(n)}(x)=(1-x)F_C^{(n)}(x)+nF_C^{(n-1)}(x)\label{ee1}\eeq
where 
 \bea F_C^{(n)}(x)&=&\frac{P_C(x)}{1-x}\sum_{S\in{\cal S}^n} C(S) \left(\frac{P_C(x)b_+}{(1-x)b_-}\right)^{h_S}\times\nonumber\\&&\quad\sum_{S'\cup S''=S}
\left(\frac{b_T}{b_+}\right)^{h_{S''}}
 H(S';P_{C_1}(x))H(S'';P_{T_1}(x)).\nonumber\\\label{ee2} \eea
It is then straightforward to generalise this formula to
\bea H(S,P_C(x))&=&{\cal R}+\left(P_C(x)\right)^{h_S}\sum_{S'\in{\cal S}^{\abs{S}}\atop{ S'\le S} } C(S,S') \left(\frac{P_C(x)b_+}{(1-x)b_-}\right)^{h_{S'}}\times\nonumber\\&&\quad\sum_{S''\cup S'''=S'}
\left(\frac{b_T}{b_+}\right)^{h_{S'''}}
 H(S'';P_{C_1}(x))H(S''';P_{T_1}(x)),\nonumber\\&&\quad
\nonumber\\ \label{master} \eea
where the leading terms are written out explicitly and ${\cal R}$ contains contributions depending only on lists inferior to ${\cal S}^{\abs S}$. Every term on the right hand side is positive so it can be bounded above by using $P_C(x)<P_\sharp(0)$ and then the expectation value taken;  moving the $S''=S$ term to the left hand side  gives 
\bea \avg{H(S,P_C(0))}\left(1-\left(\frac{P_\sharp(0)^2b_+}{b_-}\right)^{h_S}\right)\le\nn\\
{\cal R}+\left(P_\sharp(0)\right)^{h_S}\sum_{S'\in{\cal S}^{\abs{S}}\atop{ S'\le S} } C(S,S') \left(\frac{P_\sharp(0)b_+}{b_-}\right)^{h_{S'}}\times\nonumber\\\quad\sum_{{S''\cup S'''=S'}\atop S''\ne S}
\left(\frac{b_T}{b_+}\right)^{h_{S'''}}
\avg{ H(S'';P_{C_1}(0))}\avg{H(S''';P_{T}(0))}.\nonumber\\ \eea
Part (i)  is true for $\bar S=[1]$ so the lemma then follows immediately by induction on $S$. To prove part (ii) use part (i) to isolate the potentially divergent terms in (\ref{master}) leaving
\bea  H(\bar S,P_C(x))&=&
\left(\frac{P_C(x)^2b_+}{(1-x)b_-}\right)^{h_{\bar S}}
\left(
H(\bar S;P_{C_1}(x))
+\left(\frac{b_T}{b_+}\right)^{h_{\bar S}}H(\bar S;P_{T_1}(x))\right)
\nonumber\\&&\quad
+{\rm finite~terms}.\eea
For small enough $x$,
\beq 0 <\left(\frac{P_C(x)^2b_+}{(1-x)b_-}\right) < 1, \quad \forall C\eeq
and part (ii) follows upon taking expectation values.

\subsection{$\ep_1=0$, $\ep_2>0$\label{last}}

We will show that 
\beq F_S=\avgA{\frac{H(S,P_C(x))}{(1-P_C(x))^{h_S+1}}}\eeq
is finite at $x=0$, which together with (\ref{Qnderiv}) gives the result. Using
(\ref{submaster}) and (\ref{PLupper4}) gives
\beq F_S < \avgA{n_C^{h_S+1} H(S,P_C(x))},\label{inter}\eeq
where $n_C$ is the location of the first infinite tooth of $C$.
Applying (\ref{master}) iteratively we find that the right hand side is bounded above by terms of the form 
\beq \avgA{n_C^K}\avgA{H(S',P_T(x))}.\eeq
The maximum value of $K$ occurring is $h_S+1+\Phi_S$ where $\Phi_S$ is the number of strings inferior to $S$.  As remarked before $\avgA{H(S',P_T(x))}$ is analytic and $\avgA{n_C^K}$ is trivially finite which completes the proof.

\section{Calculation of $P_C(x)$ for some useful combs\label{AppL}}

Let the comb $C$ have $T_k=N_\ell, k<L$ and arbitrary $T_L$ and $C_L$. Then following  the method of Appendix A of  \cite{comb} we find
\beq  \fl{P_C^{\ep_1\ep_2}(x)= P_{\flat \ell}^{\ep_1\ep_2}(x)\left(1+
\frac{(1-A)(P_{C_{L-1}}^{\ep_1\ep_2}(x)-P_{\flat \ell}^{\ep_1\ep_2}(x))}
{A^{L-1}(P_{C_{L-1}}^{\ep_1\ep_2}(x)-AP_{\flat \ell}^{\ep_1\ep_2}(x))-(P_{C_{L-1}}^{\ep_1\ep_2}(x)-P_{\flat \ell}^{\ep_1\ep_2}(x))}\right)} \label{Ldecay}\eeq
where
\beq A=\frac{(1-x)b_-}{(P_{\flat \ell}^{\ep_1\ep_2}(x))^2b_+}.\eeq

Setting $\ep_2=0, \ep_1 <0$, $\ell=1$, $T_L=N_\infty$ and $C_L=\sharp$ we find after some algebra that
\beq P^{\ep_10}_C(x)=P^{\ep_10}_{\sharp}(x)\left(1+A^{-L}\left(x^\half\frac{A-1}{2\ep_1}+O(x)\right)\right)\label{PLupper}\eeq
and,  as $x\to 0$,
\beq A\to \frac{1+\abs{\ep_1}}{1-\abs{\ep_1}}.\eeq
Repeating the exercise but with $C_L=*$ yields a similar result.

If instead we set  $\ep_2>0, \ep_1 <0$, $\ell=1$, $T_L=N_\infty$ and $C_L=\sharp$ we find 
\beq P^{\ep_1\ep_2}_C(x)=P^{\ep_1\ep_2}_{\sharp}(x)\left(1+\frac{2\ep_2(A-1)A^{-L}}{ \ep_1-2\ep_2(1-A^{-L})}    \left( 1+ O(x)\right)\right)\label{PLupper2}\eeq
and,  as $x\to 0$,
\beq A\to \frac{1+\abs{\ep_1}-\ep_2}{1-\abs{\ep_1}-\ep_2}.\eeq
Again, repeating the exercise but with $C_L=*$ yields a similar result.

With $\ep_2>0, \ep_1 =0$, and $C=\{T_{k<L}=N_\ell, T_{k\ge L}=N_\infty\}$ we find that
\beq P^{0\ep_2}_C(x) >  P_{\flat \ell}^{\ep_1\ep_2}(x)-\frac{1}{L-1},\qquad L>2,\label{PLlower3}\eeq
(it is good enough to use $P_*(x)$ for $k=2$);
and for  $C=\{T_{k\ne L}=N_1, T_{L}=N_\ell\}$,  $x<x_0$,
\beq P^{0\ep_2}_C(x) <  P_{\sharp}^{0\ep_2}(x)\left(1-\frac{1}{
\frac{A^{L-1}-1}{A-1}+\frac{BA^{L-1}}{1-P_{N_\ell}^{\ep_2}(x)}}\label{PLupper3}\right)\eeq
where $A=(1-x) (P_{\sharp}^{0\ep_2}(x))^{-2}$ and $B$ is a positive constant depending on $x_0$, $A$ and $\ep_2$. 

Finally for $\ep_2>0, \ep_1 =0$, and $C=\{T_{k\ne L}=N_1, T_{L}=N_\infty\}$ we find that 
\beq P^{0\ep_2}_C(0) =1-\frac{1}{L+(1-\ep_2)/4\ep_2}.\label{PLupper4}\eeq

\section{Upper bound on $P(x)$ when $\ep_1=0$, $\ep_2>0$ \label{AppPc}}
We start by writing
\beq \label{hbar} H(x;n) = \avg{D_{\abs{T_n}}(x)}\avgp{\frac{G_C(x;n)}{1-P_C(x)}}, \eeq
where the measure $\bar{\mu}$ is defined in \eqref{muBar}.  Applying Jensen's inequality with this measure to \eqref{recur} results in a recursion formula of the same form as discussed in \ref{AppL} and it is easy to verify that $\avgp{P_{C_k}(x)} \ge \avg{P_{C_k}(x)}$ to give
\bea 
\avgp{\frac{G_C(x;n)}{1-P_C(x)}} &\ge& \avg{\frac{G_C(x;n)}{1-P_C(x)}} \nn \\
& \ge& \frac{3}{b_+(1-x)^{n/2}} \frac{\exp{\left(-n\avg{1-P_C(x)}\right)}}{\avg{1-P_C(x)}},
\eea
where in the last line we have again used Jensen's inequality when averaging over the ensemble.  Applying this result to \eqref{hbar}, summing over $n$, and using \eqref{unitarity} we obtain the inequality 
\bea
\frac{2}{x} \ge \avg{D_{\abs{T_n}}(x)} \frac{B}{\avg{1-P_C(x)}^2}.
\eea
Using the value for $\avg{D_{\ell}(x)}$ given in table \ref{Table1} and rearranging gives the upper bound on $P(x)$ quoted in \ref{RLUM}.


\section*{References}

\begin{thebibliography}{99}

\bibitem{one} G. H. Weiss and S. Havlin, \emph{Some properties of a random walk on a comb 
structure}, Physica \textbf{134A} (1986) 474-484
\bibitem{two} S. Revathi, V. Balakrishnan, S. Lakshmibala and K. P. N. Murthy, \emph{Validity 
of the mean-field approximation for diffusion on a random comb}, Phys. Rev. E \textbf{54} 
(1996) 2298-2302
\bibitem{three} D. ben-Avraham and S. Havlin, \emph{Diffusion and reactions in fractals and 
disordered systems}, Cambridge University Press, Cambridge (2000)
\bibitem{halv} S. Havlin, J. E. Kiefer and G. H. Weiss, \emph{Anomalous diffusion on a random 
comblike structure}, Phys. Rev. A \textbf{36} (1987) 1403-1408
\bibitem{four} J. Ambj\o rn, B. Durhuus and T. Jonsson, \emph{Quantum geometry: a statistical 
field theory approach}, Cambridge University Press, Cambridge (1997)
\bibitem{dh}J.~Ambj\o rn and Y.~Watabiki, {\it Scaling in  quantum gravity,} Nucl. Phys. {\bf B445} (1995) 129-144, hep-th/9501049
\bibitem{specdim1}J.~Ambj\o rn, J.~Jurkiewicz and R.~Loll, {\it Spectral dimension of the universe,} Phys.~Rev.~Lett. {\bf 95} (2005) 171301, hep-th/0505113
\bibitem{Pot1} C. Aslangul, P. Chvosta and N. Pottier, \emph{Analytic study of a model of diffusion on a random comblike structure}, Physica A \textbf{203} (1994) 533-565
\bibitem{six} S. Havlin, A. Bunde, H. E. Stanley and D. Movsholvitz, \emph{Diffusion on 
percolation clusters with a bias in topological space: non-universal behaviour}, J. Phys. A 
\textbf{19} (1986) L693-L698
\bibitem{eight} V. Balakrishnan and C. Van den Broeck, \emph{Transport properties on a random 
comb}, Physica A \textbf{217} (1995) 1-21
\bibitem{seven} S. Havlin, A. Bunde, Y. Glaser and H. E. Stanley, \emph{Diffusion with a 
topological bias on random structures with a power-law distribution of dangling ends}, Phys. 
Rev. A \textbf{34} (1986) 3492-3495
\bibitem{Pot3} N. Pottier, \emph{Diffusion on random comblike structures: field-induced trapping effects}, Physica A \textbf{216} (1995) 1-19 
\bibitem{Pot2} N. Pottier, \emph{Analytic study of a model of biased diffusion on a random comblike structure}, Physica A \textbf{208} (1994) 91-123 
\bibitem{comb} B. Durhuus, T. Jonsson and J. F. Wheater, \emph{Random walks on combs}, J. 
Phys. A \textbf{39} (2006) 1009-1038, hep-th/0509191
\bibitem{trees} B. Durhuus, T. Jonsson and J. F. Wheater, \emph{The spectral dimension of 
generic trees}, math-ph/0607020
\bibitem{trees2} B. Durhuus, T. Jonsson and J. F. Wheater, \emph{On the spectral dimension of 
generic trees}, DMTCS proc. \textbf{AG} (2006), 183-192
\bibitem{TanyaThesis} T.M. Elliott, Oxford University D.Phil Thesis, in preparation.
\bibitem{Fell2} W. Feller, \emph{An introduction to probability theory and its applications}, 
Vol.2, Wiley, London (1968)
\bibitem{Point} A. Grigoryan and T. Coulhon, \emph{Pointwise estimates for transition probabilities of random walks in infinite graphs}, in: Trends in mathematics: Fractals in Graz 2001, Ed. P. Grabner and W. Woess. Birkh\"aueser (2002)

\end{thebibliography}
\end{document}